\documentclass{article}
\usepackage{cite}
\usepackage{amsmath,amssymb,amsfonts}
\usepackage{algorithmic}
\usepackage{graphicx}
\usepackage{algorithm,algorithmic}
\usepackage{hyperref}
\usepackage{comment}
\usepackage{subcaption}
\usepackage{doi}
\hypersetup{hidelinks=true}
\usepackage{textcomp}
\usepackage{url}
\usepackage{adjustbox}
\usepackage{multirow}
\usepackage[symbol]{footmisc}
\providecommand{\keywords}[1]{\textbf{\textit{Index terms---}} #1}

\title{CLARAE: Clarity Preserving Reconstruction AutoEncoder for Denoising and Rhythm Classification of Intracardiac Electrograms}

\author{
  Long Lin$^{1,}$\thanks{Corresponding author (e-mail: llin@pa.uc3m.es)} \and
  Pablo Peiro-Corbacho$^{1}$ \and
  Pablo Ávila$^{2,3}$ \and
  Alejandro Carta-Bergaz$^{2,3}$ \and
  Ángel Arenal$^{2,3}$ \and
  Gonzalo R. Ríos-Muñoz$^{1,2,3,}\thanks{These authors contributed equally}$ \and
  Carlos Sevilla-Salcedo$^{4,\dagger}$
}

\date{
  $^{1}$ Department of Bioengineering, Universidad Carlos III de Madrid, Legan\'es 28911, Spain \\
  $^{2}$ Instituto de Investigación Sanitaria Gregorio Marañón (IiSGM), Madrid 28007, Spain \\
  $^{3}$ Centro de Investigación Biomédica en Red de Enfermedades Cardiovasculares (CIBERCV), Instituto de Salud Carlos III, Madrid 28029, Spain \\
  $^{4}$ Department of Signal Theory and Communications, University Carlos III of Madrid, Legan\'es, 28911 Spain
}

\begin{document}

\maketitle

\begin{abstract}

Intracavitary atrial electrograms (EGMs) provide high-resolution insights into cardiac electrophysiology but are often contaminated by noise and remain high-dimensional, limiting real-time analysis. We introduce CLARAE (CLArity-preserving Reconstruction AutoEncoder), a one dimensional encoder–decoder designed for atrial EGMs, which achieves both high-fidelity reconstruction and a compact 64-dimensional latent representation. CLARAE is designed to preserve waveform morphology, mitigate reconstruction artifacts, and produce interpretable embeddings through three principles: downsampling with pooling, a hybrid interpolation convolution upsampling path, and a bounded latent space.

We evaluated CLARAE on 495,731 EGM segments (unipolar and bipolar) from 29 patients across three rhythm types (AF, SR300, SR600). Performance was benchmarked against six state-of-the-art autoencoders using reconstruction metrics, rhythm classification, and robustness across signal-to-noise ratios from –5 to 15 dB. In downstream rhythm classification, CLARAE achieved F$_1$-scores above 0.97 for all rhythm types, and its latent space showed clear clustering by rhythm. In denoising tasks, it consistently ranked among the top performers for both unipolar and bipolar signals.

In order to promote reproducibility and enhance accessibility, we offer an interactive web-based application. This platform enables users to explore pre-trained CLARAE models, visualize the reconstructions, and compute metrics in real time. Overall, CLARAE combines robust denoising with compact, discriminative representations, offering a practical foundation for clinical workflows such as rhythm discrimination, signal quality assessment, and real-time mapping.

\end{abstract}


\section{Introduction}
\label{sec:introduction}

Intracardiac electrograms (EGMs) provide localized, high-resolution views of cardiac activation that complement the global perspective of the traditional surface electrocardiogram (ECG) and are central to arrhythmia diagnosis and therapy. Atrial fibrillation (AF) remains the most prevalent sustained arrhythmia and a growing public-health burden driven by ageing and cardio-metabolic comorbidities \cite{AF}. In this context, improving the fidelity and interpretability of EGMs is crucial for mapping, diagnosis, and treatment not only in AF workflows but across supraventricular arrhythmias such as atrial flutter and focal atrial tachycardia.

Intracavitary atrial EGMs are typically acquired with multi-electrode diagnostic catheters (e.g., circular/spiral, mini-basket, grid) in unipolar and bipolar configurations, enabling simultaneous near-field sampling at multiple sites along the cardiac endocardium. Contemporary electroanatomical mapping (EAM) systems localize the catheter, via magnetic or impedance methods, and integrate these signals into three-dimensional maps for voltage and activation annotation across supraventricular arrhythmias~\cite{Luengo2019,Rios-Munoz2022,Carta-Bergaz2025}. This acquisition technology provides rich spatiotemporal information, but it also introduces modality-specific signal characteristics and constraints that shape downstream processing.

As a result, intracardiac EGM signals are exposed to diverse artifacts with distinct temporal and spectral profiles, such as far-field interference (e.g., ventricular components within atrial recordings), baseline drift, motion/contact noise, power-line interference, and hardware-related effects. These non-stationary and frequently overlapping components can complicate conventional filtering, risking attenuation of clinically relevant near-field content or distortion of waveform morphology \cite{Tedrow2011,Nairn2020}. Robust denoising is therefore essential: cleaner EGMs improve annotation, activation-rate estimation, feature extraction, and, ultimately, the accuracy of mapping and classification across supraventricular arrhythmias.

Deep learning has produced effective denoisers for surface ECGs, e.g., deep recurrent neural networks (DRNNs) using long short-term memory (LSTM) networks to model temporal dependencies \cite{drnn}, or multi-branch convolutional approaches for baseline wander removal~\cite{deepfilter}. In parallel, convolutional denoising autoencoders (AEs) have evolved from fully convolutional deep AEs~\cite{fcn-dae} and variants with dense output layers~\cite{cnn-dae} to attention-enhanced designs~\cite{ACDAE} and fully-gated AEs with self-organized operational neural networks (self-ONNs) units~\cite{FGDAE}. However, most of these methods were developed and evaluated on ECGs, whose noise profile and morphology differ meaningfully from intracavitary EGMs, limiting their direct transfer.

Beyond denoising, compact and informative latent spaces can support interpretable machine learning and multiple downstream tasks in cardiac electrophysiology. In intracardiac settings, AEs and variational autoencoders (VAEs) have started to show promise: residual CNNs have been trained on unipolar EGMs to detect putative focal and rotational activity sources in AF~\cite{ClassificationUnipolarElectrogramsinHumanAtrialFibrillationFaST,Rios-Munoz2022CNNs}; disentangled VAEs on unipolar EGMs have been used to infer catheter positions across atrial regions~\cite{dVAE}; and denoising VAEs have demonstrated advantages over classical filtering for complex clinical electrophysiology signals such as monophasic action potential recordings~\cite{VAEIntracardiacTimeSeries}. In \emph{ECG} research, self-supervised/unsupervised AEs, e.g., masked AEs for pretraining and convolutional VAEs for large-scale representation learning, have achieved strong performance in downstream arrhythmia classification~\cite{UnsupervisedPre-TrainingUsingMaskedAutoencoders,cvae}. These results motivate AE-based latent spaces that both preserve clinically relevant information and enable robust classification across atrial rhythms.

\begin{figure*}[tbh]
\centering
\includegraphics[width=0.85\textwidth]{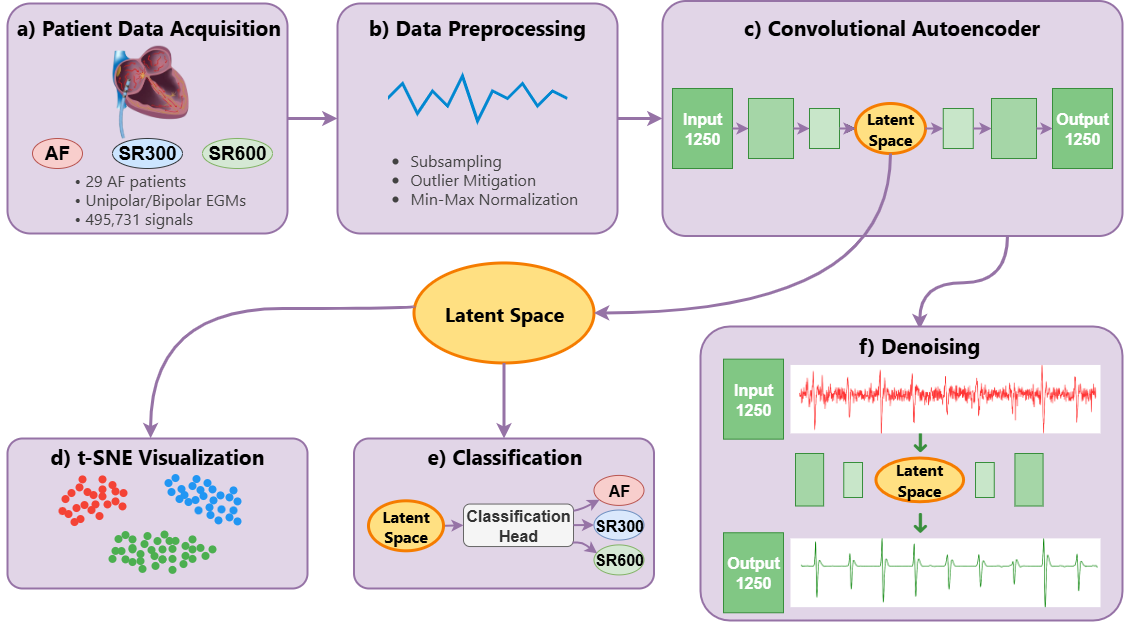}
\caption{Deep Learning Workflow for Intracardiac Electrograms Analysis. a) Patient data acquisition from atrial fibrillation (AF) patients; b) Data preprocessing; c) A convolutional autoencoder that compresses the signal into a latent space and then reconstructs it; d) Visualization of the latent space, showing separation between rhythms; e) Classification of the rhythms using features extracted from the latent space; and f) Denoising, where a noisy signal is processed to produce a clean reconstructed output.}
\label{fig:pipeline}
\end{figure*}

Previous work often treats denoising and classification separately, and generalizes from ECG to EGM without addressing intracardiac-specific artifacts (e.g., far-field contamination, catheter-tissue contact variability, or mapping-catheter geometry)~\cite{Tedrow2011,Nairn2020}. There is a need for approaches that (i) jointly optimize high-fidelity reconstruction (for denoising) and compact latent encoding (for downstream tasks), and (ii) operate consistently on both unipolar and bipolar EGMs across supraventricular arrhythmias.

In this work, we address these two key challenges in atrial EGM analysis by developing the CLArity preserving Reconstruction AutoEncoder (CLARAE), which is capable of: (1) performing dimensionality reduction through a latent representation that preserves clinically relevant information for rhythm discrimination among AF, sinus rhythm at 300\,ms (SR300), and sinus rhythm at 600\,ms (SR600); and (2) effectively denoising intracardiac atrial EGMs to improve signal quality and interpretability, independently of the specific supraventricular rhythm. Although our classification experiments use AF/SR protocols for ground evaluation, the denoising component is rhythm-agnostic by design and is intended for general application to supraventricular atrial recordings.

The remainder of this paper is structured as follows. Section~\ref{sec:methods} details the dataset, EGM acquisition, preprocessing, the proposed CLARAE model, training/metrics, and a head-to-head comparison against six baselines with latent-space and SNR-robustness protocols. Section~\ref{sec:results} reports reconstruction, rhythm classification from latent features, and denoising robustness, with qualitative examples and an interactive web demo. Sections~\ref{sec:discussion}–\ref{sec:conclusion} discuss implications/limitations and conclusions.

\section{Materials and Methods}\label{sec:methods}

We first describe the study design, data sources, and acquisition workflow for intracavitary atrial EGMs, including recording configurations. Next, we outline the preprocessing steps and provide a high-level overview of the proposed AE-based framework (illustrated in Fig.~\ref{fig:pipeline}) and state-of-the-art baseline models, followed by the training procedures and evaluation metrics used for reconstruction, representation quality, and downstream rhythm discrimination. 


\subsection{Data Collection}

Data for this study were collected from 29 patients with persistent AF who underwent an ablation procedure at the Hospital General Universitario Gregorio Marañón, Madrid (Spain). Table \ref{tab:table1} summarizes the demographic and clinical characteristics of the cohort.

\begin{table}[tbh!] 
\centering 
\caption{Database demographic and clinical statistics. These statistics were only available for 27 of the 29 patients (93.1\%). Values are mean ± std or n (\%), where percentages are calculated with respect to the number of patients for whom the statistics were available. BMI, body mass index; COPD, Chronic Obstructive Pulmonary Disease; TAI, transient ischemic attack; IHD, ischemic heart disease; CHA2DS2-VASc, Congestive Heart Failure, Hypertension, Age, Diabetes, Previous Stroke; NYHA, New York Heart Association classification; CVD, Cardiovascular Disease; LVEF, left ventricular ejection fraction; LA, Left Atrium; and LAA, Left Atrial Appendage.}
\label{tab:table1}
\begin{adjustbox}{max width=\columnwidth}
\setlength{\tabcolsep}{3pt}
\begin{tabular}{llc} \hline 
\textbf{Demographic and Clinical Data}& & \textbf{Results} \\ \hline 
Number patients & ~ & 27 (100.0) \\ 
Age (years) & ~ & 60.3 ± 9.6 \\ 
Weight (kg) & ~ & 92.7 ± 17.5 \\ 
Height (cm) & ~ & 170.6 ± 9.4 \\ 
BMI (kg/m$^2$) & ~ & 31.7 ± 4.7 \\
\textbf{Symptoms} & ~ & ~ \\
Any & ~ & 21 (77.8) \\ 
Palpitations & ~ & 11 (40.7) \\ 
Dyspnoea & ~ & 10 (37.0) \\ 
\textbf{Comorbidities} & ~ & ~ \\ 
Hypertension & ~ & 18 (66.7) \\ 
Diabetes mellitus & ~ & 5 (18.5) \\ 
Obesity & ~ & 11 (40.7) \\ 
Heart Failure & ~ & 11 (40.7) \\
Dyslipidemia & ~ & 8 (29.6) \\ 
COPD & ~ & 2 (7.4) \\ 
Obstructive sleep apnea & ~ & 4 (14.8) \\ 
TAI & ~ & 1 (3.7) \\ 
IHD & ~ & 3 (11.1) \\ 
CHA2DS2-VASc & 0 & 2 (7.4) \\ 
~ & 1 & 5 (18.5) \\ 
~ & 2 & 4 (14.8) \\ 
~ & 3 & 6 (22.2) \\ 
~ & 4 & 3 (11.1) \\ 
NYHA & I & 11 (40.7) \\ 
~ & II & 9 (33.3) \\ 
~ & III & 16 (38.1) \\ 
~ & IV & 1 (3.7) \\ 
Nº Previous CVD & ~ & 1.7 ± 1.1 \\ 
\textbf{Echocardiographic parameters} & ~ & ~ \\ 
LVEF (\%) & ~ & 53.2 ± 12.0 \\ 
LA volume (cm\textsuperscript{3}) & ~ & 142.0 ± 50.7 \\ 
LAA volume (cm\textsuperscript{3}) & ~ & 11.1 ± 6.8 \\
\hline
\end{tabular} 
    
\end{adjustbox}
\end{table}

The signals were acquired using the CARTO 3 electroanatomical mapping system (Biosense Webster, Diamond Bar, CA, USA) in conjunction with a multi-electrode PentaRay catheter (Biosense Webster). This imaging technology creates accurate 3D maps of the heart chambers by using electromagnetic fields to determine the spatial position of the catheter's electrodes. The EGMs capture local electrical activity by measuring the electrical potential through direct catheter contact with the atrial walls.

Both bipolar and unipolar EGM configurations were simultaneously obtained for each recording segment. Bipolar signals measure the potential difference between two adjacent electrodes of the catheter inside the heart. In contrast, unipolar signals measure the potential difference between an intracardiac electrode and a distant reference electrode. All signals were originally recorded at a 1 kHz sampling frequency for 2.5-second segments.

\subsection{Signal Preprocessing}
Prior to analysis, the data underwent a two-step preprocessing procedure. First, outlier mitigation was performed by clipping signals to the 0.5th and 99.5th voltage percentiles calculated from the training set to remove extreme values. Subsequently, min-max normalization was applied to scale the clipped signals to the range [-1, 1] and signals were subsampled by a factor of 2 to reduce computational cost, reducing the sampling rate from 1 kHz to 500 Hz and resulting in 1,250 samples per signal. The final dataset comprised 495,731 signals for each EGM type (unipolar and bipolar), with an average of 17,094 ± 5,881 signals per patient, distributed as follows:

\begin{itemize}
\item Atrial Fibrillation (AF): 192,531 signals (38.8\%).
\item Sinus Rhythm at 300ms (SR300): 168,539 signals (34.0\%).
\item Sinus Rhythm at 600ms (SR600): 134,661 signals (27.2\%).
\end{itemize}

We split the data into training (80\%), validation (10\%), and test (10\%) sets using patient-wise splitting to prevent data leakage. This approach ensures that signals from the same patient do not appear in multiple sets.

\subsection{CLARAE Model}

\begin{figure*}[tbh]
\centering
\includegraphics[width=0.95\textwidth]{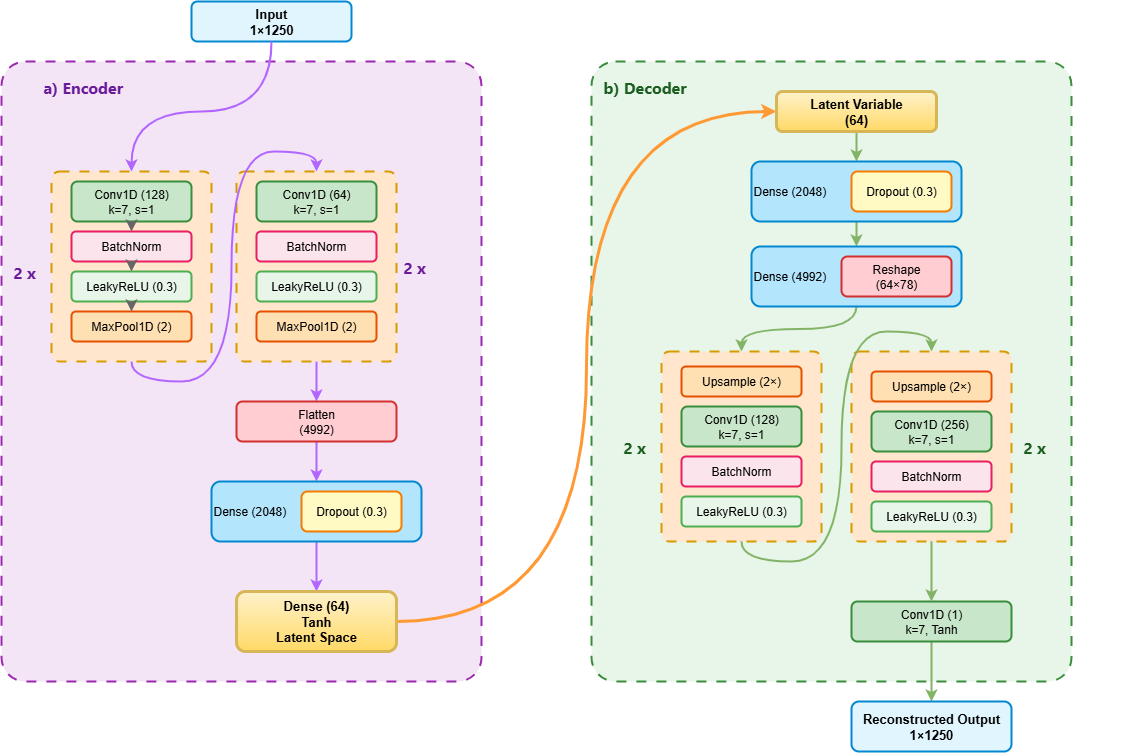}
\caption{CLARAE architecture. (a) Encoder with pooling-based downsampling to a 64-dimension bounded latent. (b) Decoder with linear-upsampling and transposed convolutions for artifact-reduced reconstruction of a \(1{\times}1250\) signal.}
\label{fig:AE_architecture}
\end{figure*}
\emph{CLARAE} is a symmetric 1D encoder–decoder customized for intracavitary atrial EGMs. Unlike prior ECG-oriented autoencoders, our model is explicitly designed to preserve clinically relevant signal features while remaining robust to common architectural pitfalls. It is built with three goals: (i) preserve near-field morphology essential for expert interpretation, (ii) produce a compact, well-structured latent representation suitable for visualization, clustering, and downstream classification tasks, and (iii) mitigate reconstruction artifacts introduced by aggressive striding or naive deconvolution.

CLARAE addresses limitations of prior deep AEs (e.g., FCN-DAE and CNN-DAE), which reduce and recover sequence length using strided or transpose convolutions \cite{cnn-dae, fcn-dae}. These operations, though sufficient for coarse ECG morphology, tend to blur temporal resolution and smear sharp deflections such as QRS complexes, features that are critical in atrial EGM interpretation. In contrast, CLARAE explicitly prioritizes sharp, artifact-free reconstructions. It does so through three key design choices that balance architectural novelty with practical robustness:
\begin{itemize}
    \item \textbf{Pooling for sharper detectors.} Explicit downsampling via MaxPool1D decouples receptive-field growth from kernel learning, preserving sharper feature detectors while controlling temporal resolution.
    \item \textbf{Hybrid upsampling path.} A two-step strategy, (1) linear interpolation and then (2) Conv1D, suppresses the “checkerboard” artifacts of naive transpose convolutions while retaining network capacity.
    \item \textbf{Bounded latent with hyperbolic tangent.} The latent code, constrained with the hyperbolic tangent activation, stabilizes the distribution and improves interpretability for downstream visualization and classification tasks.
\end{itemize}
These chore concepts are implemented in a symmetric encoder–decoder structure, described below.

\textbf{Encoder.} 
The encoder comprises two convolutional blocks followed by a compact latent head (Fig.~\ref{fig:AE_architecture}). Block~1 applies two \texttt{Conv1D} layers (128 filters, kernel size 7), each followed by batch normalization, LeakyReLU ($\alpha{=}0.3$), and \texttt{MaxPool1D} (stride 2). Block~2 mirrors Block~1 with 64-filter \texttt{Conv1D} layers. A two-layer fully-connected head, with dropout (rate 0.2) to avoid overfitting, maps the resulting sequence to a 64-dimensional latent vector (cross-validated) with $\tanh$ activation. Unlike FCN-DAE and CNN-DAE, which reduce length via strided convolutions, CLARAE uses pooling to retain sharper feature detectors while controlling temporal resolution.

\textbf{Decoder.} 
The decoder is symmetric: two dense layers expand from the 64-dimensional code, followed by two upsampling stages. 
Each stage applies \emph{linear interpolation} (factor 2) followed by a \texttt{Conv1D} layer with batch normalization and LeakyReLU. A final \texttt{Conv1D} layer with $\tanh$ activation outputs a single-channel reconstruction with the original length (Fig.~\ref{fig:AE_architecture}). 

\textbf{Complexity and compression.} 
For typical hyperparameters (e.g., $d{=}64$, input length 1{,}250), CLARAE contains $\sim$21M parameters, with over 95\% concentrated in the two fully-connected encoder/decoder layers. This design captures global morphology while leaving the convolutional path focused on local waveform structure. For 1{,}250-sample inputs, the compression ratio is $19.5{:}1$: signals are reduced to a $d$-dimensional latent representation and reconstructed at full length (Fig.~\ref{fig:AE_architecture}).

\subsection{Training Configuration and Optimization}

The model was trained using the following configuration:

\begin{itemize}
    \item \textbf{Optimizer:} Adam with an initial learning rate of 0.001, $\beta_1 = 0.9$, $\beta_2 = 0.999$.

    \item \textbf{Learning rate scheduling:} ReduceLROnPlateau with a factor of 0.5, patience of 5, and a minimum learning rate of $1 \times 10^{-8}$.

    \item \textbf{Loss function:} Mean Squared Error (MSE) for reconstruction tasks.

    \item \textbf{Batch size:} 256, selected based on GPU memory constraints and convergence stability.

    \item \textbf{Maximum epochs:} 300 with early stopping (patience = 11, $\text{min\_delta} = 1 \times 10^{-6}$).

    \item \textbf{Software:} Python 3.10, PyTorch libraries, and Weight and Biases to track training experiments.

    \item \textbf{Hardware:} NVIDIA GeForce RTX 2080 Ti with 11GB memory.
\end{itemize}

\subsection{Comparative Evaluation}

To rigorously evaluate the performance of the proposed model, it was benchmarked against several baseline architectures. This approach was chosen to situate our results within the context of current state-of-the-art methods in cardiac signal processing. The evaluation was divided into two main components: an analysis of the latent space representation and an assessment of denoising robustness.

\subsubsection{Baseline Architectures for Comparison}
To establish the improvements obtained by our proposal, six additional state-of-the-art architectures were implemented:
\begin{itemize}
    \item \textbf{Deep Recurrent Neural Network (DRNN):} An LSTM-based architecture for modeling temporal sequences, featuring a 64-unit LSTM layer, three dense layers with ReLU activation, and 30\% dropout regularization \cite{drnn}.
    \item \textbf{DeepFilter:} An architecture that employs multi-scale linear and non-linear filter modules with varying kernel sizes ([3, 5, 9, 15]) and dilated convolutions to capture features at different temporal resolutions \cite{deepfilter}.
    \item \textbf{CNN-DAE and FCN-DAE:} Two convolutional denoising autoencoder models. The CNN-DAE model, adopted in the experimentation of \cite{cnn-dae}, is very similar to FCN-DAE, except that the output transpose convolution layer is replaced by a dense (fully connected) layer \cite{cnn-dae, fcn-dae}.
    \item \textbf{ACDAE:} An attention-enhanced autoencoder that incorporates the Convolutional Block Attention Module (CBAM) \cite{ACDAE}.
    \item \textbf{FGDAE:} A fully gated denoising autoencoder that utilizes self-organizing operational neural network layers ($q=2$ configuration)~\cite{FGDAE}.
\end{itemize}

\subsubsection{Latent Space Analysis}
The quality of the features learned by the autoencoder models was assessed through two procedures to gain a detailed understanding of the latent space's utility.
DRNN and DeepFilter models were excluded from this analysis since their architecture is not autoencoder-based.
First, t-SNE visualization was used to visually inspect the separability of the rhythm classes by generating 2D projections of the 64-dimensional latent space. Second, for rhythm classification, a single-layer MLP classifier with 32 hidden units was trained on the extracted latent features to quantify their discriminative power. 

\subsubsection{Denoising Performance Evaluation}
To assess the robustness of each architecture against noise, a systematic evaluation protocol was implemented. White Gaussian noise was added to test signals at SNR levels ranging from -5 to 15~dB in 1~dB increments. 

The signal-to-noise ratio (SNR) in decibels is defined as:
\begin{equation}
\text{SNR}_{\text{dB}} = 10 \log_{10} \left( \frac{P_{\text{signal}}}{P_{\text{noise}}} \right),
\label{eq:snr}
\end{equation}
where $P_{\text{signal}}$ is the power of the clean EGM signal and $P_{\text{noise}}$ is the power of the additive white Gaussian noise.

It should be noted that the reference EGM signals inherently contain acquisition noise. Therefore, the reported SNR values reflect the ratio between the original EGM recording (considered as the reference signal) and the added Gaussian noise component.

Each of the seven models was then evaluated on its ability to reconstruct the original signal from the noisy input. Finally, performance metrics were calculated at each noise level to generate robustness curves, allowing for a direct comparison of performance across all architectures.

\subsection{Evaluation Metrics}

Model denoising performance was assessed using the \textbf{Mean Squared Error} (MSE):
\begin{equation}
    \text{MSE} = \frac{1}{N} \sum_{i=1}^{N} (y_i - \hat{y}_i)^2,
\end{equation}
where $y_i$ represents the original signal values and $\hat{y}_i$ the reconstructed signal values.

For classification tasks, the F$_1$-score was computed for each rhythm class:
\begin{equation}
F_{1} = \frac{2 \, TP}{2 \, TP + FP + FN},
\end{equation}
where $TP$ denotes the number of true positives, $FP$ the number of false positives, and $FN$ the number of false negatives.
\begin{figure}[tbp]
\centering
\begin{subfigure}[t]{0.48\textwidth}
    \centering
    \includegraphics[width=\textwidth]{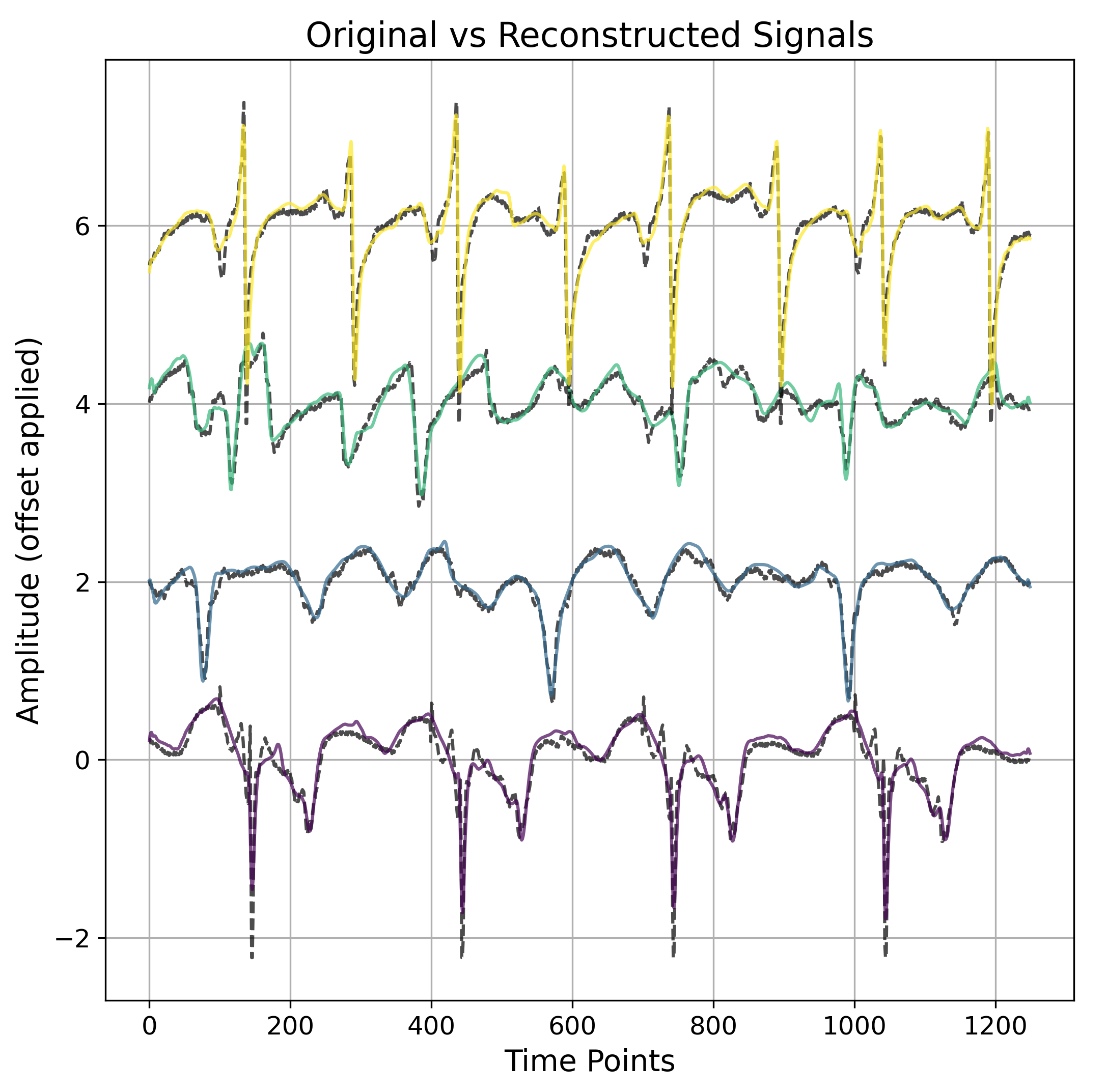}
    \caption{Unipolar EGMs. Original (dotted) vs. reconstructed (solid).}    \label{fig:latent_representation_unipolar}
\end{subfigure}
~
\begin{subfigure}[t]{0.48\textwidth}
    \centering
    \includegraphics[width=\textwidth]{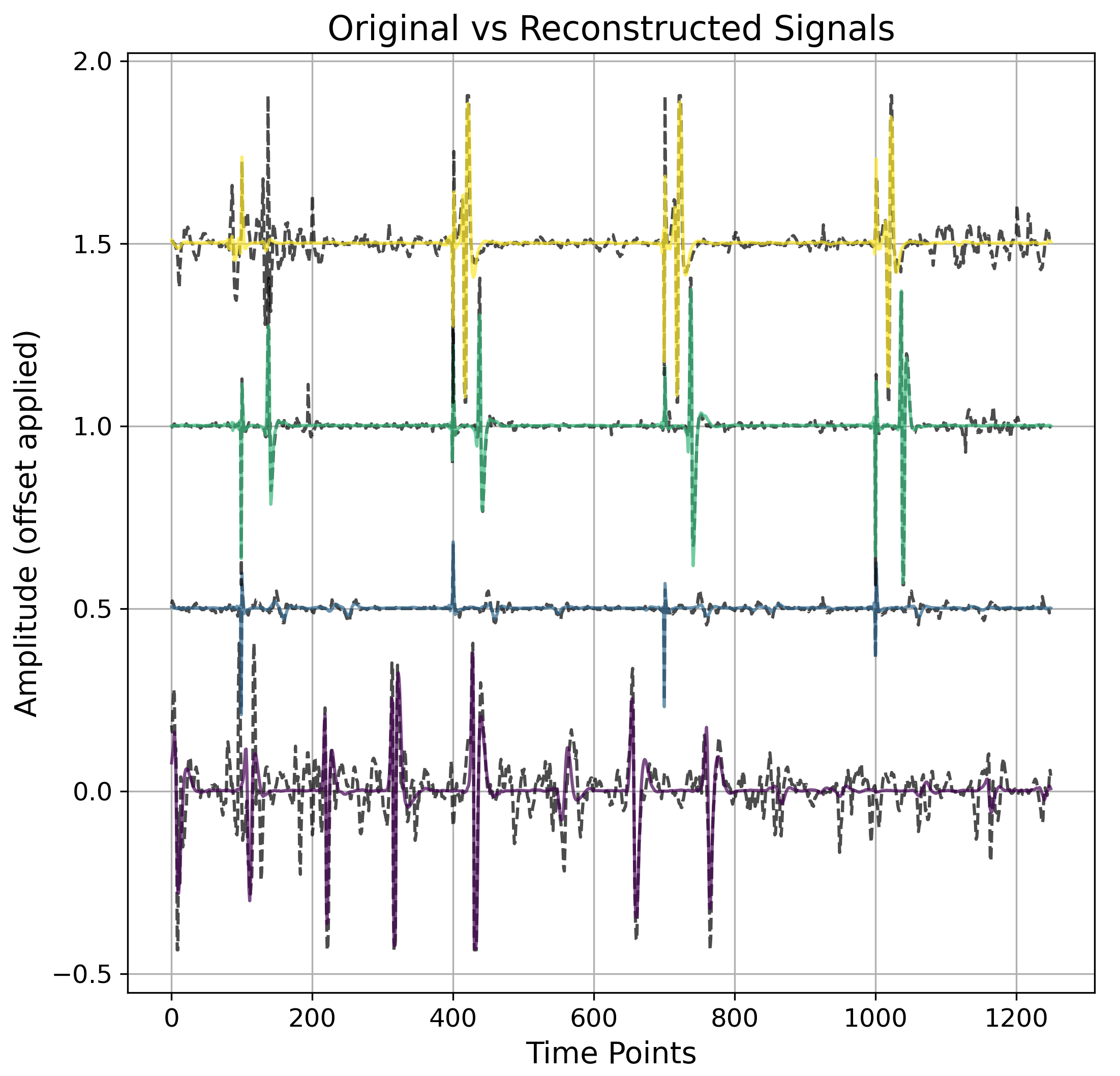}
    \caption{Bipolar EGMs. Original (dotted) vs. reconstructed (solid).}   \label{fig:latent_representation_unipolar}
    \label{fig:latent_representation_bipolar}
\end{subfigure}

\caption{CLARAE performance on EGM reconstruction. 
Across both (a) unipolar and (b) bipolar EGMs, the autoencoder preserves morphology while compressing signals by 95\% (from 1,250 to 64 values).} 
\label{fig:latent_representation_combined}
\end{figure}

\subsection{Interactive Web Application}

To facilitate reproducibility and broader accessibility of our approach, we developed an interactive web application using the Python Dash framework. The application allows users to: 
\begin{enumerate}
    \item Upload pre-trained CLARAE model weights in \texttt{.pth} format.
    \item Upload unipolar or bipolar EGM signals in text or spreadsheet format.
\end{enumerate}

Upon processing, the application generates the denoised signal and computes the Mean Squared Error (MSE) relative to the input. The source code is openly available at \url{https://github.com/longlin20/CLARAE}. This tool provides a reproducible and user-friendly platform for exploring CLARAE’s denoising performance and latent space representations.

\section{Results}\label{sec:results}

\subsection{EGM Reconstruction Performance}

Because intracavitary EGMs are primarily interpreted through their waveform morphology, faithful signal reconstruction is a prerequisite for any latent representation to be clinically meaningful. We therefore benchmarked CLARAE against six established baselines in terms of MSE on both unipolar and bipolar EGMs. Results are summarized in Table~\ref{tab:reconstruction_performance}. CLARAE achieved a test MSE of 0.011 for unipolar and 0.008 for bipolar signals.

 
\begin{table}[hbt]
\centering
\caption{Reconstruction performance in terms of MSE of CLARAE and baseline models on unipolar and bipolar EGMs. Best results per column are highlighted in bold.}
\label{tab:reconstruction_performance}
\renewcommand{\arraystretch}{1.2}
\begin{tabular}{lcccccccc}
\hline
\textbf{Model} & \textbf{Unipolar EGMs} & \textbf{Bipolar EGMs} \\
\hline
DRNN       & \textbf{0.001}  & \textbf{0.001} \\
DeepFilter & \textbf{0.001} & \textbf{0.001} \\
CNN\_DAE   & 0.020  & 0.017 \\
FCN\_DAE   & 0.027 & 0.016 \\
ACDAE      & \textbf{0.001}  & \textbf{0.001} \\
FGDAE      & 0.003  & \textbf{0.001} \\
CLARAE     & 0.011 & 0.008 \\
\hline
\end{tabular}
\end{table}

Representative reconstructions obtained by our proposal are shown in Fig.~\ref{fig:latent_representation_combined}. In both unipolar and bipolar EGMs, CLARAE reproduces the near-field morphology without introducing visible artifacts, while compressing each 1,250-sample signal into a 64-dimensional latent vector.

\begin{figure*}[tbp]
\centering
\includegraphics[width=0.88\textwidth]{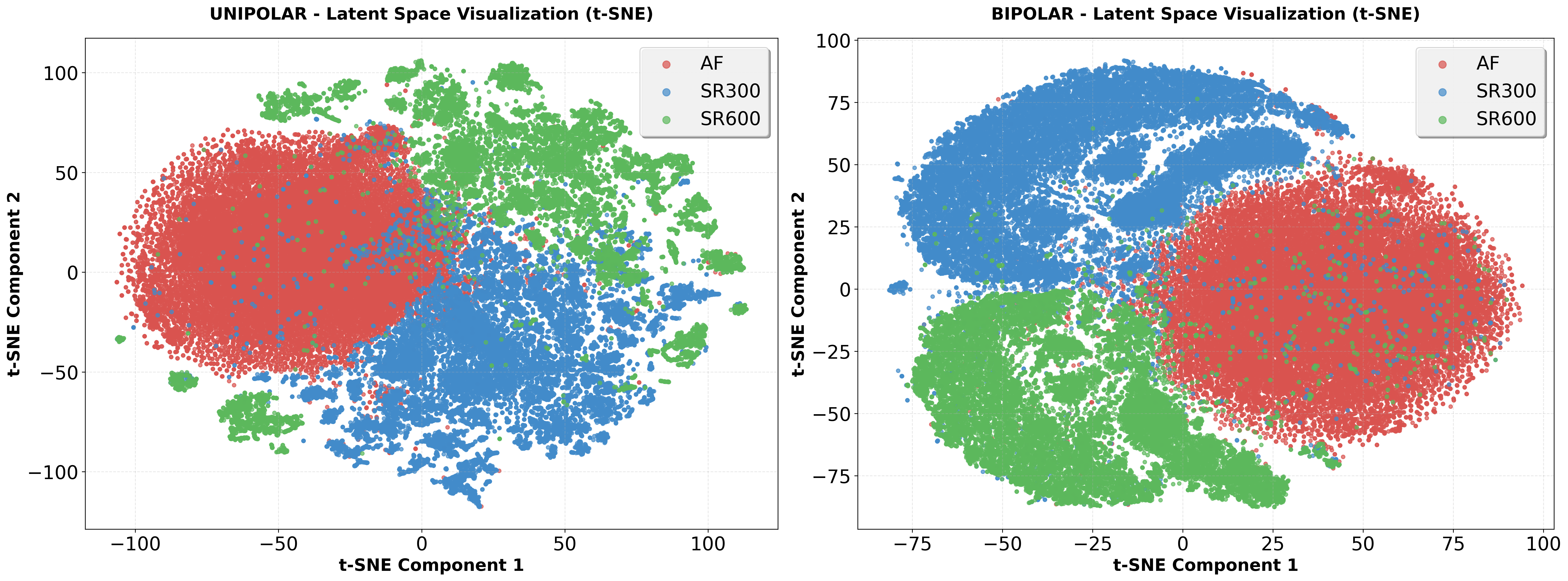}
\caption{t-SNE visualization of the 64-dimensional latent space for unipolar (left) and bipolar (right) EGMs. Clear separation between AF (red), SR300 (blue), and SR600 (green) rhythm types demonstrates the effectiveness of the learned representation.}
\label{fig:tsne_visualization}
\end{figure*}

\subsection{Latent Space Analysis and Classification Results}

Beyond reconstruction fidelity, an effective autoencoder must produce a latent space that captures task-relevant structure. To assess this, we analyzed CLARAE’s embeddings in terms of clustering and downstream classification performance.

The t-SNE visualization of the latent space (Fig.~\ref{fig:tsne_visualization}) revealed a clear separation between different rhythm types. The most striking result that emerges from the data is that AF, SR300, and SR600 signals formed distinct clusters with minimal overlap.
This observation may support the hypothesis that the 64-dimensional latent representation effectively captures the underlying electrophysiological characteristics that distinguish between different cardiac rhythms.

Using these embeddings as input to a single-layer MLP classifier achieved excellent performance in distinguishing AF, SR300, and SR600 rhythms (Table~\ref{tab:f1_scores}). CLARAE achieved the highest F$_1$-scores for all unipolar rhythms (AF: 0.981, SR300: 0.967, SR600: 0.975). While FGDAE achieved slightly higher scores on bipolar signals (AF: 0.995, SR300: 0.993, SR600: 0.991), CLARAE consistently delivered strong performance across both modalities (AF: 0.988, SR300: 0.984, SR600: 0.980), demonstrating the robustness of its latent representation. ACDAE’s embeddings were unsuitable for classification, yielding F$_1$-scores of 0 for SR300 and SR600, and are therefore excluded from the table.

\begin{table}[hbt]
\centering
\caption{Comparison of F$_1$-Scores for rhythm classification (AF, SR300, SR600) using latent features from different autoencoder models on unipolar and bipolar EGMs. Best results per column are highlighted in bold. }
\addtolength{\tabcolsep}{-0.4em}
\renewcommand{\arraystretch}{1.2}
\begin{tabular}{lcccccccc}
\hline
\multirow{2}{*}{\textbf{Model}} & \multicolumn{3}{c}{\textbf{Unipolar EGMs}} & \multicolumn{3}{c}{\textbf{Bipolar EGMs}} \\
\cline{2-4} \cline{5-7}
& AF & SR300 & SR600 & AF & SR300 & SR600 \\
\hline
CNN\_DAE & 0.969 & 0.943 & 0.956 & 0.914 & 0.931 & 0.916 \\
FCN\_DAE & 0.956 & 0.918 & 0.942 & 0.964 & 0.951 & 0.945 \\
FGDAE & 0.978 & 0.957 & 0.952 & \textbf{0.995} & \textbf{0.993} & \textbf{0.991} \\
CLARAE & \textbf{0.981} & \textbf{0.967} & \textbf{0.975} & 0.988 & 0.984 & 0.980 \\
\hline
\end{tabular}
\label{tab:f1_scores}
\end{table}


\subsection{Comparative Analysis of Denoising Performance}

\begin{figure}[tbp]
\centering
\includegraphics[width=\columnwidth]{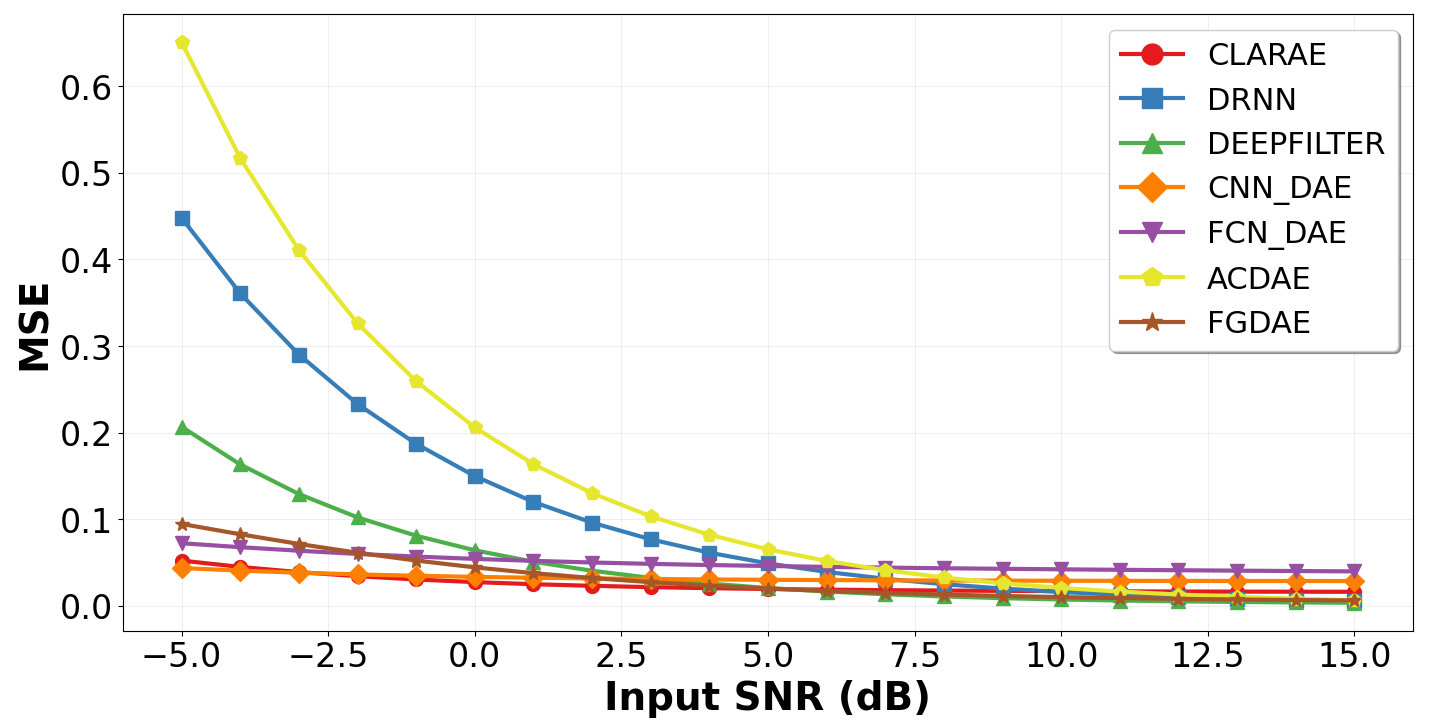}
\caption{Mean squared error (MSE) for unipolar EGMs with the different explored models at varying input signal-to-noise ratios (SNR). CLARAE demonstrates robust denoising performance across the full SNR range from -5 to 15 dB.}
\label{fig:mse_comparison}
\end{figure}

We examined the denoising capabilities of CLARAE relative to six baseline architectures under varying input noise conditions. Quantitative results are summarized in Fig.~\ref{fig:mse_comparison}, which shows the MSE as a function of input SNR for unipolar signals. As expected, all models improve with increasing SNR. CLARAE consistently ranks among the top-performing models across the full SNR range from -5 to 15 dB, closely following other leading convolutional architectures (FGDAE, FCN\_DAE, CNN\_DAE), while significantly outperforming DRNN and DeepFilter. Equivalent trends were observed for bipolar signals, confirming the robustness of these findings across both lead configurations.

\begin{figure*}[tbp]
\centering
\begin{subfigure}[t]{\columnwidth}
    \centering
    \includegraphics[width=\textwidth]{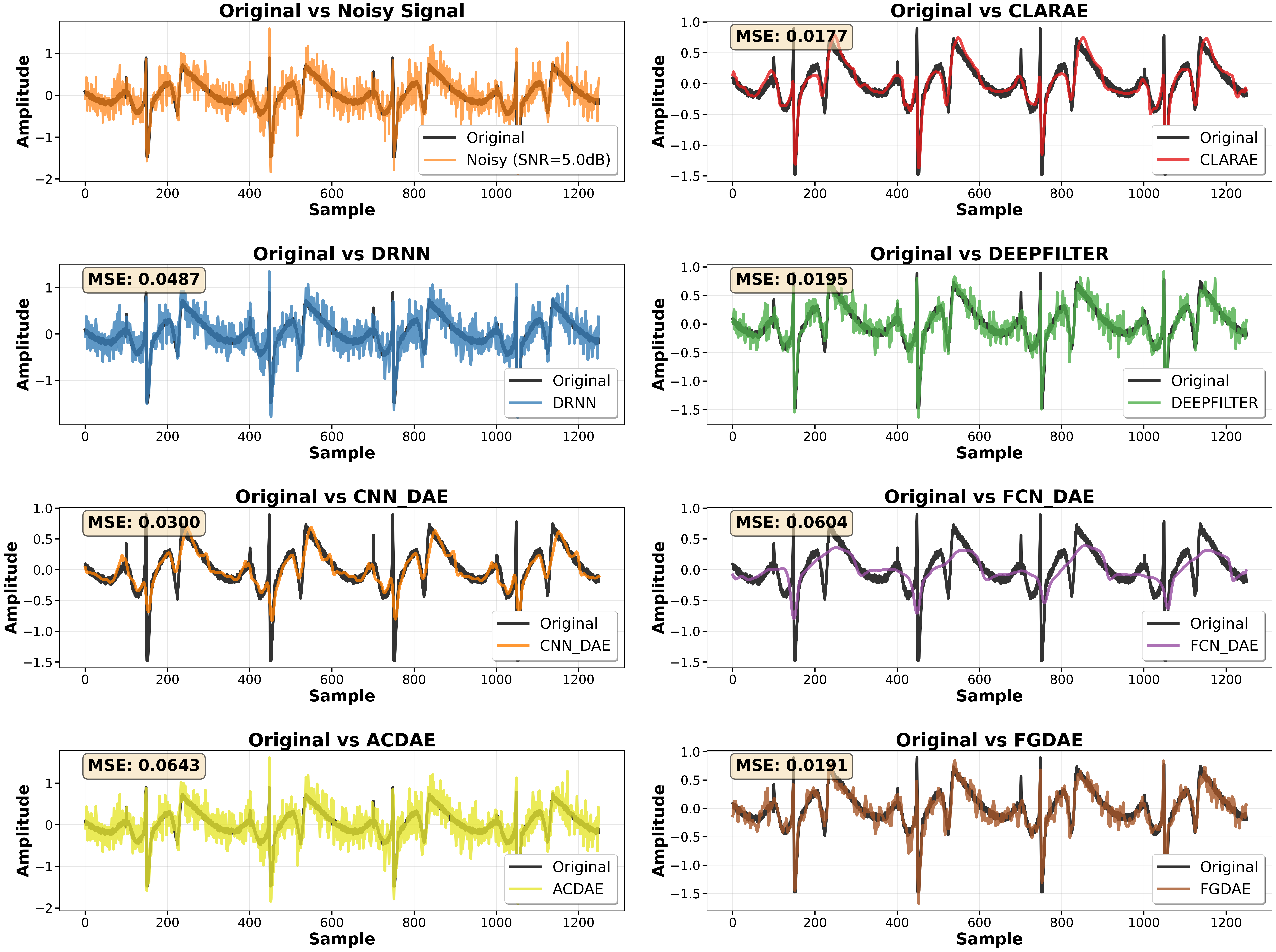}
    \caption{Unipolar EGMs.}    \label{fig:comparative_analysis_unipolar}
\end{subfigure}
~
\begin{subfigure}[t]{\columnwidth}
    \centering
    \includegraphics[width=\textwidth]{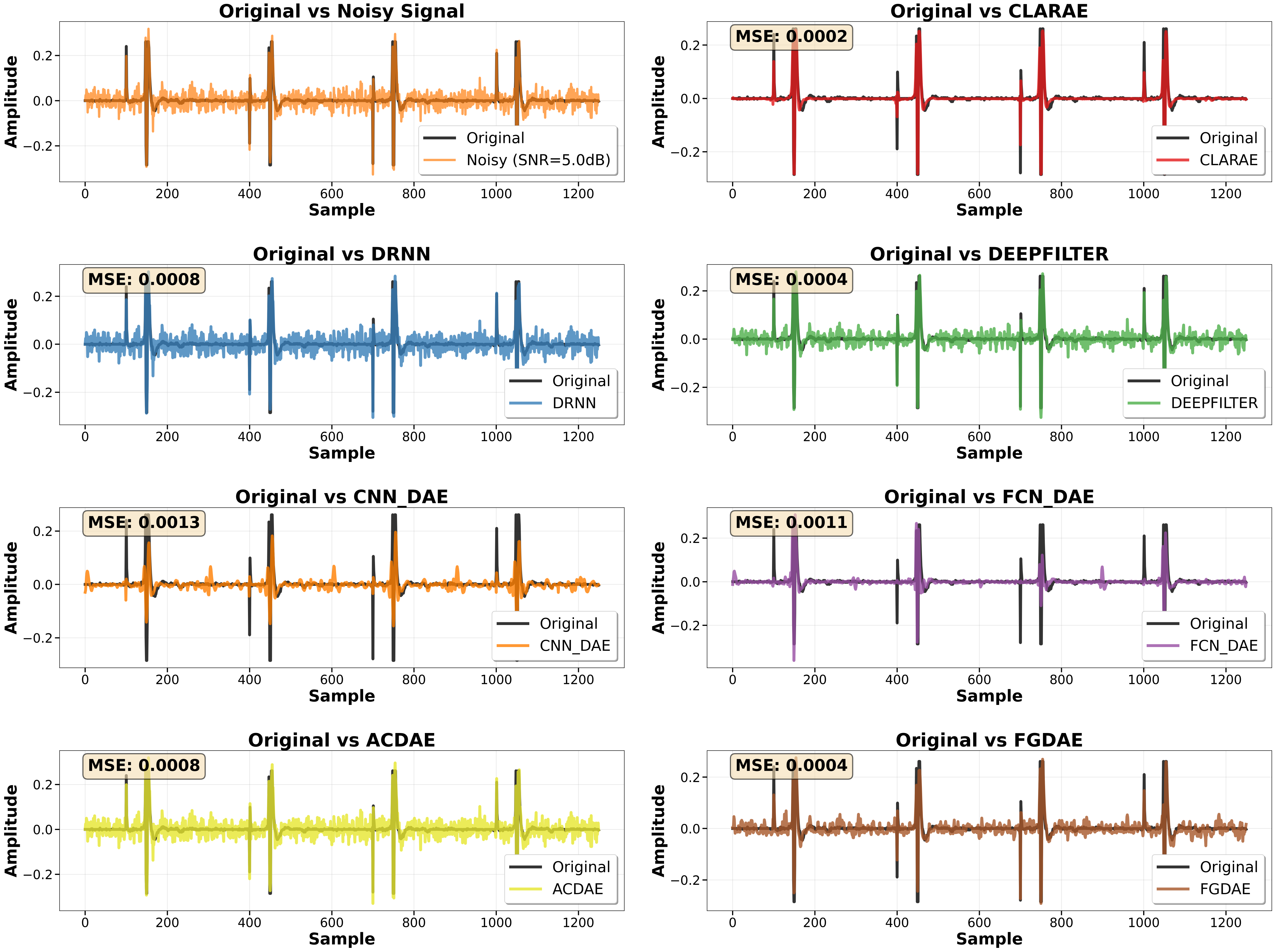}
    \caption{Bipolar EGMs.}   \label{fig:latent_representation_unipolar}
    \label{fig:comparative_analysis_bipolar}
\end{subfigure}
\caption{Example reconstructions of EGM signals at 5 dB input SNR for the different analysed models. CLARAE achieves the lowest reconstruction error while preserving signal morphology.} 
\label{fig:comparative_analysis_combined}
\end{figure*}

Representative reconstructions at 5 dB SNR are shown in Figs.~\ref{fig:comparative_analysis_unipolar} and \ref{fig:comparative_analysis_bipolar} for unipolar and bipolar signals, respectively. At this SNR, CLARAE achieved the lowest MSE among all models (unipolar: 0.0183, bipolar: 0.0002), visually demonstrating its ability to preserve signal morphology even in noisy conditions.

\subsection{Interactive Web Demonstration}

\begin{figure*}[tbh]
\centering
\includegraphics[width=\textwidth]{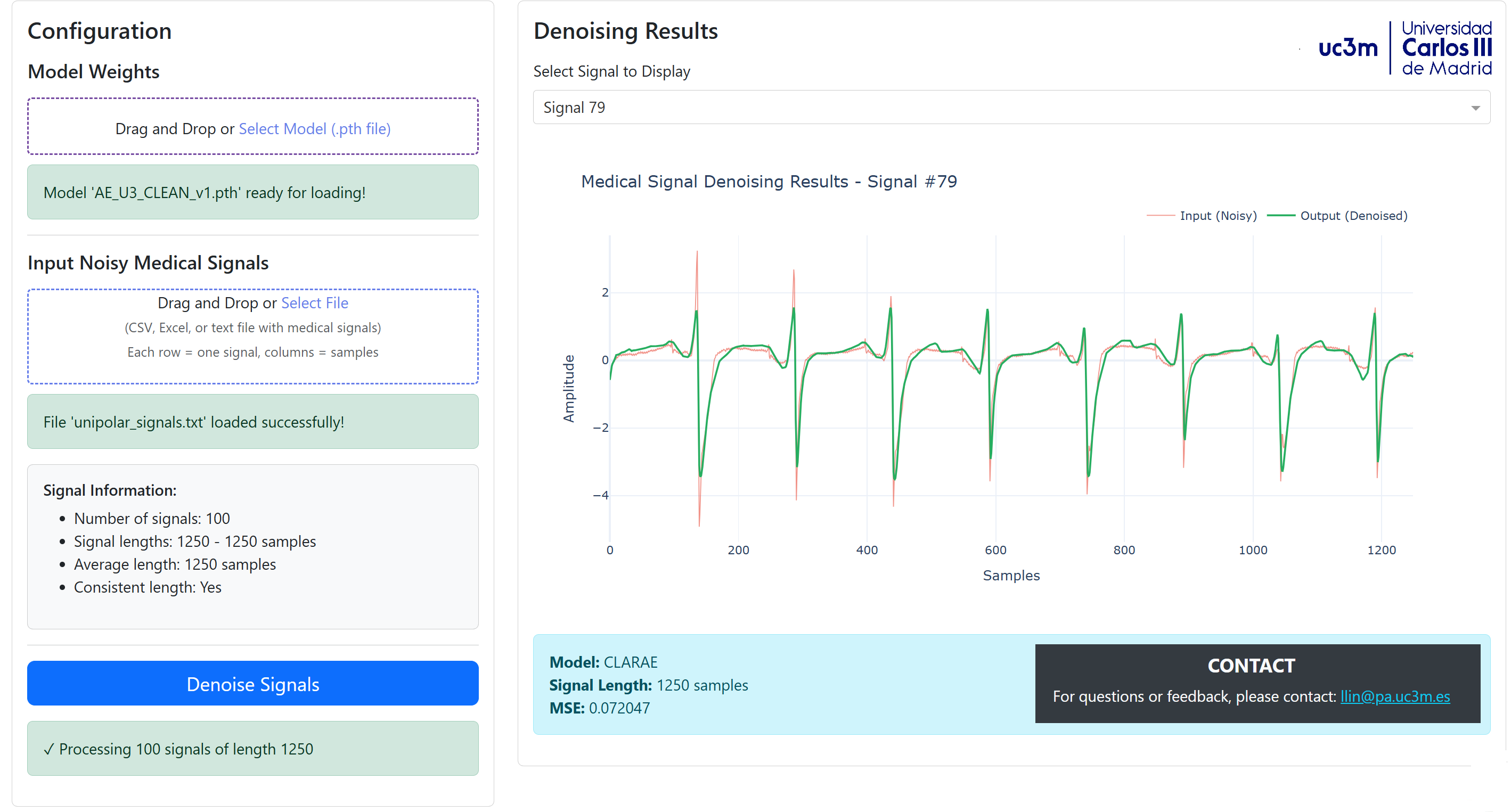}
\caption{Interactive web demonstration for EGM denoising. The interface allows users to upload pre-trained model weights (\texttt{.pth}) and their own unipolar or bipolar signals (text/spreadsheet), then visualizes the original noisy input alongside the denoised output and reports the Mean Squared Error (MSE) for immediate visual and quantitative feedback.}
\label{fig:Dash}
\end{figure*}

To illustrate the usability of CLARAE, we provide a web-based interactive demonstration (Fig.~\ref{fig:Dash}). Users can upload their own pre-trained model weights and signals to visualize both the original noisy input and the denoised output, along with the corresponding MSE. This immediate visual and quantitative feedback highlights the accessibility and effectiveness of the proposed model in practical scenarios.



\section{Discussion}\label{sec:discussion}

\subsection{Summary of Findings}
We show that CLARAE can compress intracavitary atrial EGMs from 1{,}250 samples to a 64-dimensional code (\(\approx95\%\) reduction). Despite this compression, it preserves sufficient morphology for accurate reconstruction and rhythm discrimination. Latent-space visualizations display clear clustering, and a lightweight MLP on the 64-D codes exceeds 0.97 F$_1$-score for AF, SR300, and SR600. Across unipolar/bipolar signals and SNR \(-5\) to \(15\)\,dB, CLARAE consistently ranks among the top performers for denoising and reconstruction. To the best of our knowledge, these results position compact EGM representations as practical embeddings for downstream clinical tasks. 

\subsection{What is New and Why it Matters}
CLARAE’s advantage originates from three design choices: (\emph{i}) pooling-based downsampling to decouple receptive-field growth from kernel learning; (\emph{ii}) hybrid upsampling (linear resize \(\rightarrow\) transposed convolution) to mitigate checkerboard artifacts; and (\emph{iii}) a bounded latent with \(\tanh\) that regularizes the code distribution for stable visualization/classification. These architectural choices directly translate into improved denoising, faithful reconstruction, and discriminative latent embeddings. Prior work in ECG denoising and self-supervised representation learning (e.g., GAN-style approaches \cite{Singh2021_GAN_denoise,Wang2022_cGAN_denoise}) underscores the value of architectures that suppress characteristic artifacts while retaining discriminative content; we observe analogous benefits in intracardiac EGMs, where far-field contamination and contact variability are prominent \cite{Friedman2025_LatentECG}.

\subsection{Comparative Performance and Mechanisms}

Qualitative examples in unipolar (Fig.~\ref{fig:comparative_analysis_unipolar}) and bipolar (Fig.~\ref{fig:comparative_analysis_bipolar}) traces show CLARAE suppresses interference while preserving near-field deflections. CNN-DAE and FCN-DAE remove noise but exhibit higher reconstruction error, consistent with strided-convolution downsampling blurring sharp activations; while DRNN, DeepFilter, ACDAE, and FGDAE tend to retain baseline wander and intermittent high-frequency residue at lower SNRs. Quantitative results in Table~\ref{tab:reconstruction_performance} further illustrate this trade-off: CLARAE does not achieve the lowest reconstruction error, as recurrent models such as LSTM- and RNN-based architectures often reconstruct the input nearly perfectly. However, these models also reproduce the inherent noise of the signal, effectively overfitting to artifacts. By contrast, as highlighted in Figs.~\ref{fig:comparative_analysis_unipolar} and \ref{fig:comparative_analysis_bipolar}, CLARAE prioritizes reconstructing the clean underlying morphology while discarding noise, yielding clinically more meaningful denoising despite a modest increase in error metrics.
The reconstruction results in Table~\ref{tab:reconstruction_performance} also support this finding, since CLARAE is not the best method for reconstructing the original (and noisy) signals during unsupervised training.
Our upsampling strategy is motivated by established analyses showing that pure transposed convolutions induce checkerboard artifacts, while resize\(\rightarrow\)conv mitigates them, findings that translate well to one-dimensional biomedical signals where high-frequency artifacts are clinically undesirable \cite{odena2016deconvolution}.

\subsection{Dimensionality Reduction and Downstream Utility}
The 64-dimension latent space (19.5:1 compression) retains features relevant to rhythm discrimination while remaining small enough for near–real-time use.
Beyond classification, such representations could power automated quality indices, confidence overlays on electroanatomical maps, and rapid retrieval/clustering of similar EGM segments during mapping.
The direction is aligned with recent evidence that unsupervised/SSL latent features in cardiology capture broad disease-relevant variance and support multiple downstream tasks \cite{Friedman2025_LatentECG}.

\subsection{Clinical and Engineering Implications}
An EGM denoiser that generalizes across unipolar/bipolar configurations and SNR regimes can smooth mapping sessions by presenting cleaner tracings with quantifiable quality metrics without requiring aggressive analogue filtering.
Importantly, EAM systems (Carto/EnSite/Rhythmia) increasingly rely on high-density multipolar catheters and timing heuristics (e.g., maximum negative unipolar dV/dt \cite{Rios-Munoz2018}) to reduce far-field timing errors; improved denoising and compact codes may enhance map reliability and operator workflow without adding burden \cite{Narayan2024,Rios-Munoz2022CNNs}. Beyond visual denoising, the embeddings could inform automated feature extraction, segment selection, or rhythm classification in real time, enhancing procedural safety and efficiency.

\subsection{Limitations}
First, this is a single-center dataset; performance across vendors, catheters, and signal-conditioning chains requires multi-center validation. Second, while denoising is rhythm-agnostic by design, classification experiments target AF/SR; broader supraventricular arrhythmias (flutter, focal AT) should be evaluated explicitly. Finally, we did not profile end-to-end latency on acquisition consoles; clinical deployment will require runtime benchmarks and possibly distillation and/or pruning.

\section{Conclussions}\label{sec:conclusion}

We introduce CLARAE, a 1D encoder–decoder tailored to intracavitary atrial EGMs, that simultaneously achieves high-fidelity reconstruction and a compact, discriminative 64-dimensional latent representation. It outperforms or matches existing ECG-oriented baselines across unipolar and bipolar signals over a wide SNR range, while preserving waveform morphology critical for downstream tasks such as rhythm classification.

The 64-dimensional latent space supports both visualization and efficient downstream analysis, enabling near–real-time applications in automated quality assessment, mapping, and rhythm discrimination. By combining robust denoising with interpretable embeddings, CLARAE establishes a practical foundation for using compact EGM features in clinical workflows.

\section*{Acknowledgment}
This work was supported by the Instituto de Salud Carlos III (Madrid, Spain) (PI22-01619), and by the Madrid Government (Comunidad de Madrid, Spain) under the Multiannual Agreement with UC3M (FLAMA-CM-UC3M), and through project MAGERIT-CM (TEC-2024/COM-44). 

This work involved human subjects in its research. Approval of all ethical and experimental procedures and protocols was granted by the Institutional Review Board of Hospital General Universitario Gregorio Marañón (Code: Hybrid-AFMAP, Approval date: 22nd November 2021).
\bibliographystyle{abbrv}
\bibliography{bibliography}

\end{document}